\documentclass[twocolumn,prc,aps,floatfix]{revtex4}
\usepackage[dvips]{graphicx}
\usepackage{hhline}
\begin{document}

\newcommand{\be}{\begin{eqnarray}}
\newcommand{\ee}{\end{eqnarray}}
\title{
Cosmological variation of  deuteron binding energy, strong interaction
and quark masses  from big bang nucleosynthesis}

\author{  V.F. Dmitriev$^{1,2}$,  V.V. Flambaum$^1$, and J.K. Webb$^1$ }
\affiliation{$^1$
 School of Physics, The University of New South Wales, Sydney NSW
2052,
Australia
}
\affiliation{$^2$
Budker Institute of Nuclear Physics, 630090, Novosibirsk-90,
Russia
}

\date{\today}
\begin{abstract}
We use Big Bang Nucleosynthesis calculations and light element
abundance data to constrain the relative variation of the deuteron
binding energy since the universe was a few minutes old, $\delta Q =
Q(BBN)-Q(present)$.  Two approaches are used, first treating the
baryon to photon ratio, $\eta$, as a free parameter, but with the
additional freedom of varying $\delta Q$, and second using the WMAP
value of $\eta$ and solving only for $\delta Q$.  Including varying
$Q$ yields a better fit to the observational data than imposing the
present day value, rectifying the discrepancy between the $^4He$
abundance and the deuterium and $^7Li$ abundances, {\it and} yields
good agreement with the independently determined $\eta_{WMAP}$.  The
minimal deviation consistent with the data is significant at about the
4-$\sigma$ level; $\delta Q/Q= -0.019 \pm 0.005$.  If the primordial
$^4$He abundance lies towards the low end of values in the literature,
this deviation is even larger and more statistically significant.
Taking the light element abundance data at face-value, our result may
be interpreted as variation of the dimensionless ratio
$X=m_s/\Lambda_{QCD}$ of the strange quark mass and strong scale:
$\delta X/X=(1.1 \pm 0.3) \times 10^{-3}$.  These results provide a
strong motivation for a more thorough exploration of the potential
systematic errors in the light element abundance data.
\end{abstract}
\maketitle
\vspace{0.1in}
Pacs numbers: 26.35.+c, 21.10.Dr, 98.80.Ft
\section{Introduction}
Recent astronomical data suggest a possible variation of the fine
structure constant $\alpha=e^2/\hbar c$ at the $10^{-5}$ level over a
time-scale of 10 billion years, see \cite{alpha} (a discussion of
other limits can be found in Ref. \cite{uzan} and references therein).
Naturally, these data motivated more general discussions of possible
variations of other constants. Unlike for the electroweak forces for
the strong interaction, there is generally no direct relation between
the coupling constants and observable quantitites. In recent papers
\cite{FS,DF,FS1}, we presented general discussions on the possible
influence of the strong scale variation on primordial Big Bang
Nucleosynthesis (BBN) yields, the Oklo natural nuclear reactor, quasar
absorption spectra and atomic clocks. Here we continue this work,
concentrating on BBN.

One can only measure variations of dimensionless parameters.  Big Bang
Nucleosynthes is sensitive to a number of fundamental dimensionless
parameters including the fine structure constant $\alpha$ ,
$\Lambda_{QCD}/M_{Plank}$ and $m_q/\Lambda_{QCD}$ where $m_q$ is the
quark mass and $\Lambda_{QCD}$ is the strong scale determined by a
position of the pole in the perturbative QCD runnung coupling
constant. In this work we search for any possible variation of
$m_q/\Lambda_{QCD}$ because there is a mechanism which provides a very
strong sensitivity of BBN to this parameter.

The first and most crucial step in BBN is the process $p+n\rightarrow
d+\gamma$. The synthesis starts at $t\ge 3$ sec. when the temperature
goes down below $T \le 0.6$ MeV and lasts until $t \le 6$ min. when
the temperature becomes $T \le 0.05$ MeV.  The reaction rate for the
above process defines all subsequent processes and final primordial
abundances of light elements.  Amongst the factors that can influence
the reaction rate, the most significant seems to be a variation of the
deuteron binding energy (this variation was discussed 
in Refs. \cite{Dys71,Davies,Barrow,PP91,FS,FS1,Kneller,Yoo}). Indeed, the 
equilibrium concentration of
deuterons and the inverse reaction rate depend exponentially on it.
Moreover, the deuteron is a shallow bound level.  Therefore the
relative variation of the deuteron binding $Q$ is much larger than the
relative variation of the strong potential $U$, i.e. $\delta Q/Q >>
\delta U/U$.  As a result the
variations in the strong interaction may be most pronounced via the
deuteron binding energy.
 We also take into account the effect of variation of the
virtual level in the neutron-proton system, which is even more
sensitive to the variation of the strong interaction. 

The question we address here is whether or not existing observations
of the primordial abundances of the light elements suggest any change
in the deuteron binding energy at the time of BBN.  

To do so, we use a compliation of light element abundance data from
the literature for $^4$He, $^7$Li and D/H.  As we show later, the
currently greater experimental precision on $^4$He results in that
element dominating our results.  The other 2 light elements
nevertheless provide important consistency checks.  

The data we use for $^4$He is presented in Table \ref{4he} and
comprised 14 surveys giving estimates for the primordial value, Y$_p$,
derived using, or by extrapolation to, low metallicity in each case.
There is clear evidence for significant scatter amongst these 14
values, presumably due to unquantified systematics, or if not,
intrinsic inhomogeneities.  The dominance by $^4$He, or indeed by any
single element, unfortunately increases susceptibility to systematic
errors, and we have therefore attempted to explore the effect of these
in several ways.

Firstly, in order to make best use of all the available $^4$He data,
we add a constant term to each of the statistical errors on Y$_p$,
such that that the normalised $\chi^2$ for all 14 points about the
weighted mean value is equal to unity.  This approach is equivalent to
the assumption that all 14 estimates of Y$_p$ are unbiased and
Gaussian distributed, but that there is an additional systematic
component to the statistical error which is different (and hence
random) for each estimate.

Second, as shown later, smaller values of Y$_p$ are less consistent
with $\delta Q/Q = 0$ than larger values.  Thus we carry out a
re-analysis using a subset of the Y$_p$'s, taking only the highest
values such that the normalised $\chi^2$ about the weighted mean value
is equal to unity, {\it without} increasing the individual errors by a
constant, as described above.  This procedure selects 9 values from
the original 14.  In doing this, we are exploring the consequence of
there being strong systematics for the small Y$_p$'s, and little or
none for the high values.  This is conservative, in the sense that we
are minimising our estimate for $\delta Q/Q$.  

Finally, in order to obtain some estimate of the plausible range on
our estimate of $\delta Q/Q$, we perform the converse analysis,
subsetting the data by discarding {\it high} values of Y$_p$, again
such that the normalised $\chi^2$ about the weighted mean value is
equal to unity. This leaves 9 points.  The two samples thus overlap.

\begin{table}
\caption{Data on the primordial $^4$He mass fraction}
\begin{ruledtabular}
\begin{tabular} {c|l}   \label{4he}
$Y_p$ & Ref. \\
\hline
 0.2391 $\pm$ 0.0020& \cite{lppc}  \\
 0.2384 $\pm$ 0.0025& \cite{ppl} \\
0.2371  $\pm$ 0.0015& \cite{pp}  \\
0.2443  $\pm$ 0.0015& \cite{ti}  \\
0.2351  $\pm$ 0.0022& \cite{ppl1} \\
0.2345   $\pm$ 0.0026& \cite{ppr} \\
0.244   $\pm$ 0.002 & \cite{ti1} \\
0.243   $\pm$ 0.003&  \cite{itl}  \\
0.232   $\pm$ 0.003& \cite{os} \\
0.240   $\pm$ 0.005& \cite{itl1} \\
0.234 $\pm$ 0.002 & \cite{oss} \\
0.244 $\pm$ 0.002 & \cite{it} \\
0.242 $\pm$ 0.009 & \cite{it1} \\
0.2421$\pm$ 0.0021& \cite{izth03}\\
\end{tabular}
\end{ruledtabular}
\end{table}

The data on deuterium abundances D/H from quasar absorption systems were
selected according two criteria:  \\
(i) Metallicity must be low, so as to more closely reflect primordial value:
[Si/H] or [O/H] less than or equal to -2.0. \\
(ii) Must be detection, not upper limit. \\
These requirements leave only five data points listed in Table \ref{ddata}
 \begin{table}
\caption{Data on the primordial deuterium abundance}
\begin{ruledtabular} 
\begin{tabular} {c|c|c|c|c}   \label{ddata}
QSO & z(abs)& D/H$\times 10^{-5}$& [Si/H]& Ref.\\
\hline
Q1009+299& 2.504& 4.0 $\pm$ 0.65 & -2.53&  \cite{bt} \\
PKS1937-1009& 3.572& 3.25$\pm$ 0.3& -2.26 [O/H]& \cite{bt1} \\
HS0105+1619 & 2.536& 2.5 $\pm$ 0.25& -2.0   & \cite{omear} \\
Q2206-0199   & 2.076 & 1.65 $\pm$ 0.35 & -2.23 & \cite{petb} \\
Q1243+3047 &  2.526 & 2.42 +0.35 - 0.25& -2.77 [O/H]& \cite{kirk}       \\
\end{tabular}
\end{ruledtabular}
\end{table}

The data for Lithium primordial abundance are shown in Table \ref{lidata}.
Here $A=Log(Y_{Li})+12.$
\begin{table}
\caption{Data on the primordial Li/H abundance}
\begin{ruledtabular}
\begin{tabular} {c|c}   \label{lidata}
$A$ & Ref. \\
\hline
 2.09 +0.11-0.12& \cite{rbofn}  \\
 2.35 $\pm$ 0.1& \cite{boni} \\
2.36 $\pm$ 0.12 & \cite{bdsk}  \\
2.34 $\pm$ 0.056$\pm$0.06 & \cite{bonal}  \\
2.07 + 0.16 - 0.04& \cite{syb} \\
2.22 $\pm$ 0.20& \cite{thor} \\
2.4   $\pm$ 0.2 & \cite{pswn} \\
2.5   $\pm$ 0.1&  \cite{tv}  \\
\end{tabular}
\end{ruledtabular}
\end{table}

Applying the first procedure described above, in order to obtain $\chi^2/N=1$,
we have to add to the individual $\sigma$'s 0.0017 for helium points, 
$0.344\times 10^{-5}$ for deuterium points, and 0.028 for lithium points.
For the weighted mean values we obtain
\begin{equation}  \label{hedata}
Y_p= 0.2393 \pm 0.0011,
\end{equation}
\begin{equation} \label{meand}
       Y_D=(2.63 \pm 0.31)\times 10^{-5},
\end{equation}
and
\begin{equation} \label{ali}
A = 2.315 \pm 0.051.
\end{equation}
The latter value  corresponds to the following lithium abundance
\begin{equation} \label{yli}
Y_{Li}  = (2.02 \pm 0.22 )\times 10^{-10}.
\end{equation}

The second and the third procedures are meaningful only for the helium points.
The number of deuterium points is too small and the lithium data points are 
the least scattered. We need only 20\% increase in individual uncertainties to
bring $\chi^2/N$ to 1 for the lithium data.
In addition, the deuterium and the lithium data do not produce a significant 
contribution in determination of $\delta Q/Q$ which is entirely dominated by 
the helium data due to their high accuracy.

Keeping 9 upper points for the helium mass fraction data, that give 
$\chi^2/N=0.94$, we obtain for the weighted mean value
\begin{equation} \label{hehigh}
Y_p = 0.2424 \pm 0.0008.
\end{equation}
If we keep 9 lower points, we obtain
\begin{equation} \label{helow}
Y_p = 0.2363 \pm 0.0008,
\end{equation}
which is significantly lower than both in Eq.(\ref{hehigh}) and Eq.
(\ref{hedata}).

\section{The BBN equations}
We use the standard BBN set of equations that describe the time development of
the abundances of the elements in an expanding Universe \cite{wago}
\begin{equation} \label{bbneq}
\frac{\dot{R}}{R}=H=\sqrt{\frac{8\pi}{3M_P^2}\rho_T},
\end{equation}
\begin{equation} \label{bden}
\frac{\dot{n_B}}{n_B} =-3H,
\end{equation}
\begin{equation}
\dot{\rho_T} = -3H (\rho_T + p_T),
\end{equation}
\begin{equation} \label{abun}
\dot{Y}_i = \sum_{j,k,l} N_i\left( \Gamma_{kl\rightarrow ij} \frac{Y_l^{N_l}
Y_k^{N_k}}{N_l!N_k!} - \Gamma_{ij\rightarrow kl}\frac{Y_i^{N_i}Y_j^{N_j}}
{N_i!N_j!}\right),
\end{equation}
\begin{equation} \label{neutr}
n_- - n_+ = \frac{n_B}{T^3}\sum_j Z_j Y_j,
\end{equation}
where $n_B$ is the density of baryons, $Y_i$ is the abundance of the
element $^{A_i}Z_i$. The right-hand side of Eq.(\ref{abun}) corresponds to a
reaction
\begin{equation}
N_i(^{A_i}Z_i) + N_j(^{A_j}Z_j)\leftrightarrow N_k(^{A_k}Z_k) + N_l(^{A_l}Z_l).
\end{equation}
$\rho_T$ and $p_T$ denote total energy density and pressure, respectively,
\begin{equation}
\rho_T = \rho_\gamma + \rho_e + \rho_\nu + \rho_B,
\end{equation}
\begin{equation}
p_T = p_\gamma + p_e + p_\nu + p_B.
\end{equation}
Eq.(\ref{bbneq}) defines the expansion rate. Eq.(\ref{bden}) defines
the change in time of the baryon density, and the rate equation Eq.(\ref{abun})
defines the time evolution of the abundances and their final values after
freeze-out. The last Eq.(\ref{neutr}), where $n_-$ and $n_+$ are the densities
of electrons and positrons, is the condition of electro-neutrality that defines
a chemical potential of electrons.
\section{Effect of the deuteron binding energy variation}
The sensitivity of the reaction rate $\Gamma_{\gamma d \leftrightarrow p n}$ to 
parameters of the strong interaction in general, and to the deuteron binding energy
in particular, comes from two sources. First,
the reaction rate $\Gamma_{\gamma d \leftrightarrow p n}$ depends exponentially
on the deuteron binding energy $Q$. Second, the cross section of the reaction 
 $n p \rightarrow\Gamma_{\gamma d }$ is very sensitive to the position of the virtual
level with the energy $\epsilon_v=0.07 MeV$. Any change in the strong NN-potential causing 
a shift in the deuteron binding energy $Q$ will change the position of the virtual level 
 $\epsilon_v$ as well. The relation between $\delta Q$ and $\delta \epsilon_v$ 
can be obtained using the fact that both a real level and a virtual one are 
close to $E=0$. The relation is (see Appendix) 
\begin{equation}\label{rel1}
\frac{\delta \epsilon_v}{\sqrt{\epsilon_v}}=-\frac{\delta Q}{\sqrt{Q}}.
\end{equation}
The cross section for $n+p \rightarrow d+\gamma$ reaction can be found in 
textbooks \cite{seg}. In the leading order in $Q/\epsilon_v$ the product of 
the cross section and the velocity is proportional to
$$
\sigma v \sim Q^{5/2}/\epsilon_v.
$$
 Thus, in linear order in $\delta Q$ we have the following modification
of the reaction rate
\begin{equation}\label{dgam}
\Gamma_{np\rightarrow d\gamma}\rightarrow \Gamma_{np\rightarrow d\gamma}\left(
1+ \left(5/2+\sqrt{\frac{Q}{\epsilon_v}}\right)\frac{\delta Q}{Q}\right).
\end{equation}
We should note, however, that according to our BBN calculations the direct
effect of the deuteron binding energy variation
 (due to the exponential dependence of
the inverse reaction rate $ d\gamma\rightarrow np$) is more important than
the variation of the cross-section $np\rightarrow d\gamma$ Eq.(\ref{dgam}) .
 
We modified one of the standard BBN codes \cite{kawano} in such a way
that $Q$ can be changed for this reaction.  Varying $Q$ changes the
abundances of all three elements under discussion.  In Fig.1 we plot
the abundance of D, the mass fraction of $^4$He, and the abundance of
$^7$Li, as functions of $Q$ at the value of the baryon to photon ratio
$\eta=6.14\times 10^{-10}$ found from anisotropy of cosmic microwave
background \cite{benn}
\begin{figure}
\includegraphics[width=8 cm]{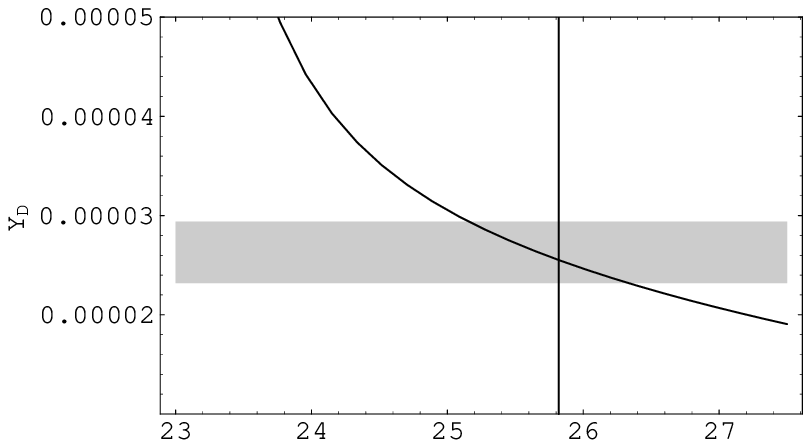}
\includegraphics[width=80 mm]{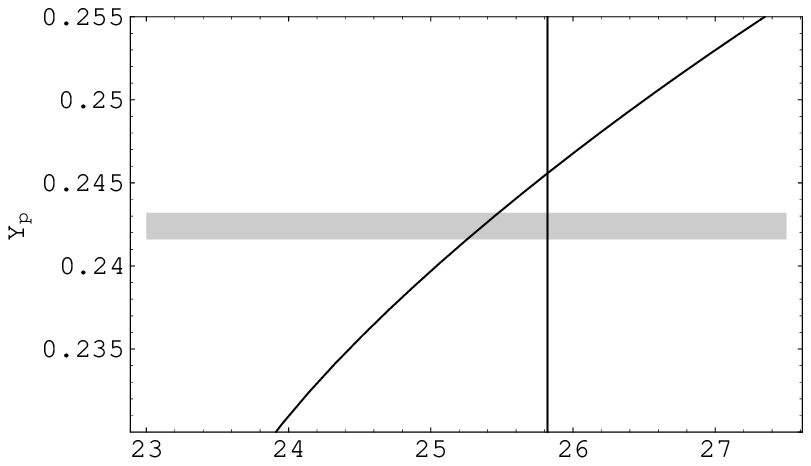}
\includegraphics[width=80 mm]{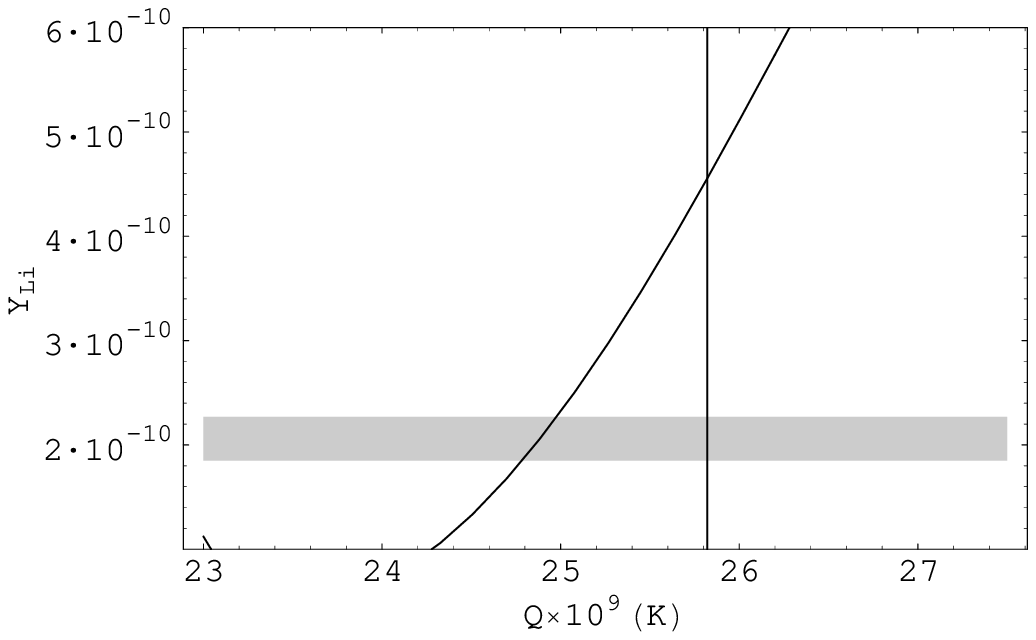}
\caption {The light elements abundances as a functions of the deuteron
binding energy $Q$ for $\eta_{WMAP} = 6.14\times 10^{-10}$.  The
vertical line shows the present value of $Q = 25.82\times 10^9$ K.
The shaded regions illustrate the 1$\sigma$ range in the data. For
helium the high $Y_p$ value (Eq.\ref{hehigh}) is shown.}
\end{figure}
From Fig 1. we see that the deuterium abundance is not very sensitive to $Q$.
The data are fully compatible with the present value of the deuteron binding
energy. Such a poor sensitivity can be explained by relatively large error bars
for the deuterium abundance.

  The data on $^4$He, in contrast, show strong sensitivity to the
deuteron binding energy favoring for lower $Q$ during primordial
nucleosynthesis.  The data on $^7$Li also favoring for lower $Q$ approximately
for the same $\delta Q$ as $^4$He.  

The above Figure 1 give a qualitative picture of the dependence of light 
element abundances on the deuteron binding energy. 
In order to obtain more
quantitative results we analyse the likelihood functions as functions of $Q$
and $\eta$.
\subsection{The likelihood functions}
 The likelihood function for the abundances have been choosen in the form
\begin{eqnarray} \label{lf}
 Lf(\eta, Q) =
\exp(-\frac{1}{2}\sum_{ij}(Y^{th}_i(\eta,Q)-Y^{ex}_i)w_{ij} \nonumber \\
 \times (Y^{th}_j(\eta,Q)-Y^{ex}_j)).
\end{eqnarray}
Here the sum goes over three light elements, $w_{ij}$ is the inverse error
 matrix that was calculated using the approach proposed in Ref.
\cite{flsv}. The errors in theoretical values of the abundances can be found
from the uncertainties in the reaction rates
 \begin{equation} \label{corr}
 \delta Y_i^{th} = Y^{th}_i\sum_k \lambda_{ik}\frac{\Delta R_k}{R_k},
\end{equation}
where $\Delta R_k$ are the reaction rate errors, and
$$
\lambda_{ik} =\frac{\partial\, {\rm ln}Y^{th}_i}{\partial\, {\rm ln}R_k}
$$
are the logarithmic derivatives. The error  matrix $\sigma_{ij}$ can be calculated
then by
\begin{equation} \label{err}
\sigma_{ij}^2 =  Y^{th}_i   Y^{th}_j\sum_k
\lambda_{ik}\lambda_{jk}\left(\frac{\Delta R_k}{R_k}\right)^2.
\end{equation}
The uncertainties in the experimental data (\ref{hedata}),
(\ref{meand}), (\ref{yli}) should be added to the diagonal matrix elements of
the error matrix (\ref{err})
\begin{equation} \label{sig}
\sigma^{tot\,2}_i = \sigma^2_{ii} + \sigma^{ex\,2}_i.
\end{equation}
For $^4$He $\sigma^{tot}$ differs from $\sigma^{ex}$ insignificantly, while for
D and especially for $^7$Li $\sigma_{ii}$ and   $\sigma^{ex}_i$ are comparable.
If we neglect the correlations then the matrix
$w_{ij}$ is diagonal and equal to $$
w_{ii} = 1/\sigma^{tot\,2}_i.
$$
In this case we can present the likelihood function (\ref{lf}) as a product of
three individual functions $Lf(\eta,Q)=
Lf_D(\eta,Q)Lf_{He}(\eta,Q)Lf_{Li}(\eta,Q) $.  The equations
 \begin{equation} \label{lines}
 Y_i(\eta,Q) = Y^{ex}_i
 \end{equation}
defines three lines in $\eta-Q$ plane where the individual likelihood functions
are equal to one.  And the equations
 \begin{equation} \label{vall}
 (Y_i(\eta,Q) - Y^{ex}_i)^2= \sigma^{tot\,2}_i
 \end{equation}
define 1$\sigma$ ranges around these lines for each element. These ranges are
shown in Fig 2. 
 \begin{figure}
\includegraphics[width=8cm]{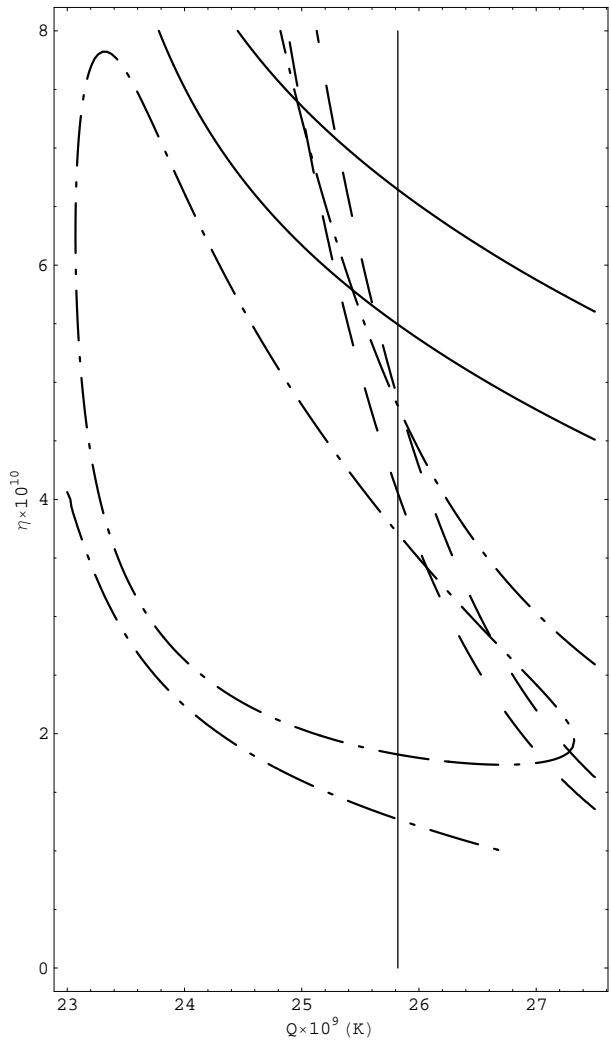}
\caption{1$\sigma$-ranges around the maxima of individual likelihood
functions.  The solid lines show 1$\sigma$-ranges for D, the dashed
lines are for $^4$He (using $Y_p$ from Eq.\ref{hehigh}), and the dot-dashed 
lines
are for $^7$Li. For lithium, there are 2 solutions for $\eta$ and $Q$,
hence the shape of the error contours is more complicated.}
\end{figure}
The slope of the deuterium range is smaller than that of helium and lithium
reflecting smaller sensitivity in $Q$ and higher sensitivity in $\eta$. .

In contrast, the helium range goes almost vertically reflecting  high 
sensitivity of the helium fraction to
$Q$ and low sensitivity to $\eta$. This low sensitivity to $\eta$ can be 
explained by a large helium binding energy. Only gamma's with the energy
$E_\gamma >$20 MeV can significantly change the number of helium nuclei. At any
$\eta$ the number of such $\gamma$-quanta is small at the BBN temperature. We
can, therefore, expect the low sensitivity of the helium mass fraction to 
$\eta$.

The Lithium range has two distinct branches
corresponding to two different solutions of Eq.(\ref{lines}) for $\eta$ at 
given $Q$.
All three ranges intersect near $\eta=6.5$ and
$Q=25$. One can expect that  the general 
likelihood
function (\ref{lf}) will have a maximum in this region. Indeed, we found the
maximum of $Lf(\eta,Q)$ at the point $\eta_m=(6.51\,+\,0.77\,-\,0.66 )\times 
10^{-10}$ and
$Q_m=(25.26\pm 0.20)\times 10^9 $K.  Fig. 3 shows 1$\sigma$ elliptic boundary 
near the maximum. The long axis of the ellipsis is almost vertical. Therefore, 
\begin{figure}
\includegraphics[width=80mm]{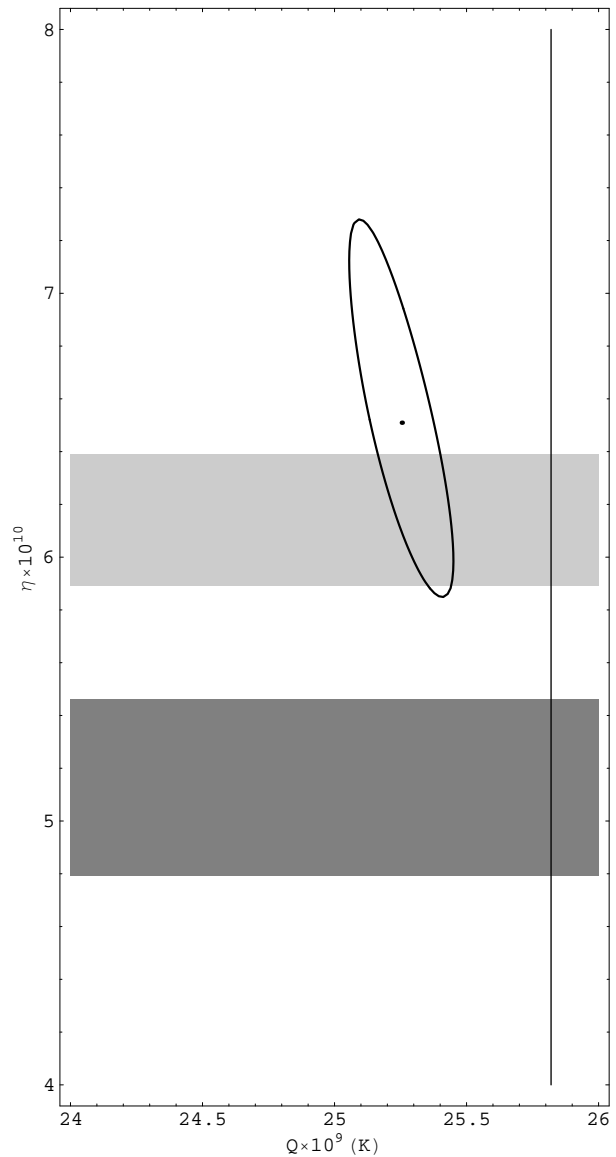}
\caption{1$\sigma$-range about the maximum of $Lf(\eta,Q)$ (again
using Y$_p$ from Eq.\ref{hehigh}). The lighter shaded region shows
CMB-WMAP data for $\eta$.  The darker shaded region is the
1$\sigma$-range for $\eta$ from BBN calculations using the present-day
value of the deuteron binding energy, $Q=25.82$.  A lower value of
Y$_p$ will produce a larger deviation between the $\eta_{WMAP}$ and
$\eta_{BBN}$.}
\end{figure}
the correlation between $\Delta\eta$ and $\Delta Q$ is not significant.
Comparing Fig. 3 and Fig. 2 one can conclude that the error   $\Delta Q$ is
determined mostly by $^4$He mass fraction data. It is interesting to note that
$\eta_m$ is compatible with the one found from recent CMB anisotropy 
measurement\cite{benn}. The dark shadow region shows the 1$\sigma$ range 
for $\eta$ fitted from BBN only at present value of $Q=25.82$ K.

\subsection{Constraint from CMB anisotropy measurements}
The value of $\eta$ found from CMB  anisotropy measurements
$$
\eta_0=(6.14\pm 0.25)\times 10^{-10}
$$
has rather high accuracy. It is natural to use the constraint from this
measurement in our study of the deuteron binding energy effects. To do this we 
construct
another likelihood function which is a function of $Q$ only.
 \begin{equation}  \label{lf0}
L(Q)=\int_{-\infty}^\infty\exp(-\frac{(\eta-\eta_0)^2}{2\sigma^2_\eta})
Lf(\eta,Q)\,d\eta.
\end{equation}
If we neglect nondiagonal elements in $w_{ij}$ we can construct the
individual likelihood functions for D, $^4$He, and $^7$Li. They are
constructed in the same way as (\ref{lf0}) using instead of general
function $Lf(\eta,Q)$ the individual ones $Lf_D(\eta,Q)$,
$Lf_{He}(\eta,Q)$ , $Lf_{Li}(\eta,Q)$.  These functions are plotted in
Fig. 4 together with the general likelihood function (\ref{lf0})
\begin{figure}
\includegraphics[width=75 mm]{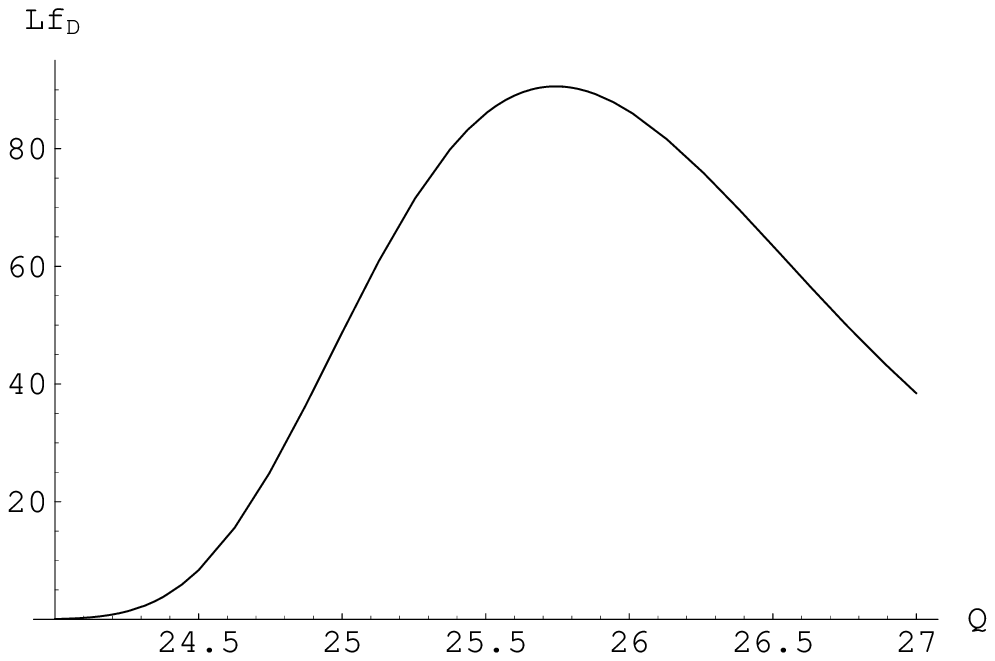}
\includegraphics[width=75 mm]{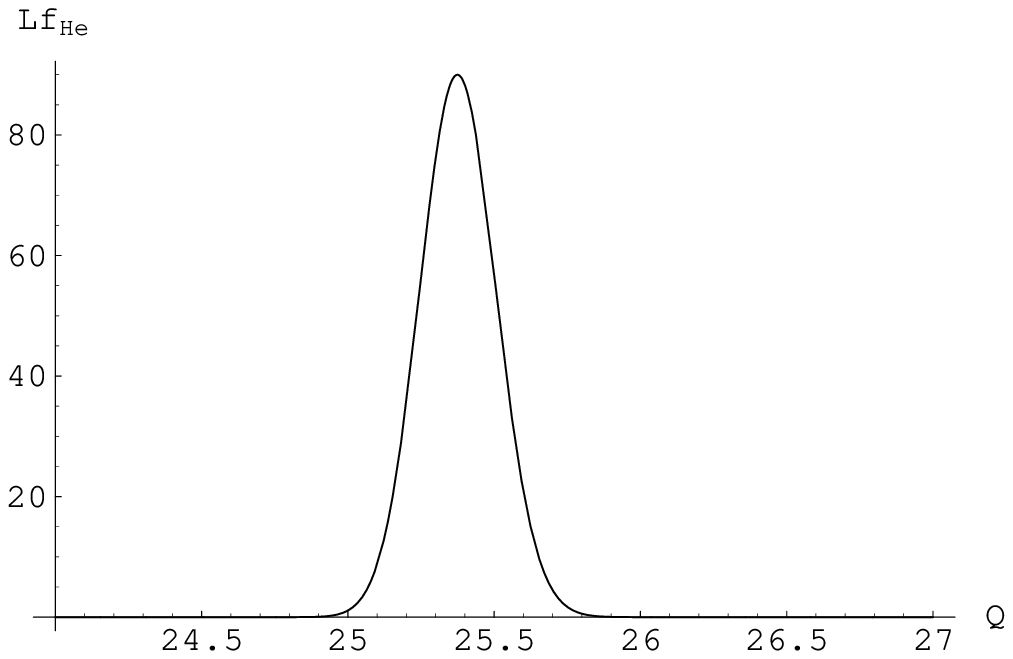}
\includegraphics[width=75 mm]{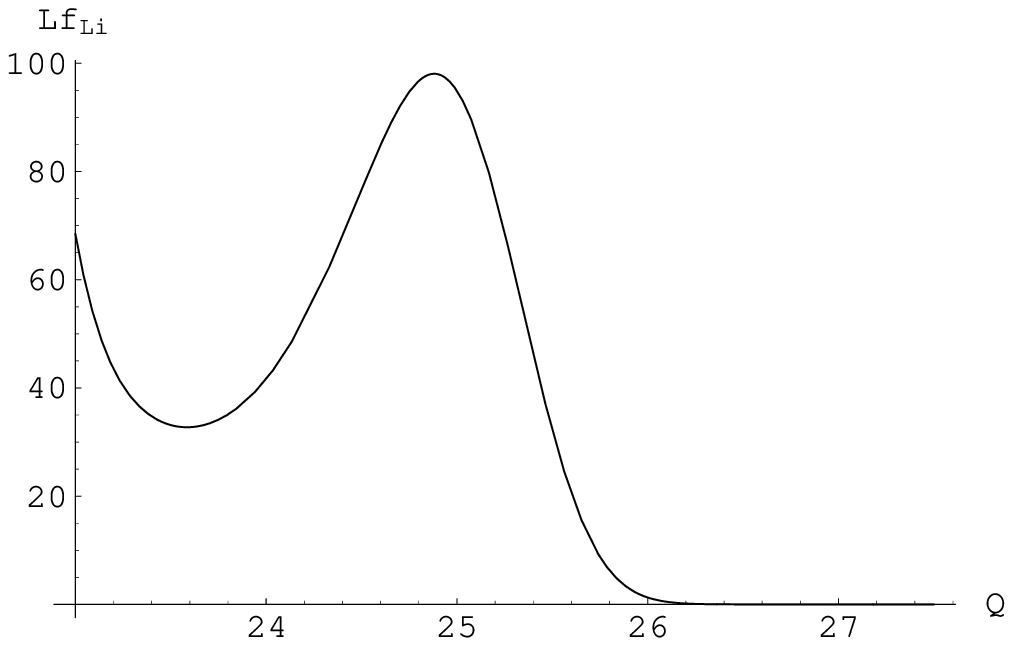}
\includegraphics[width=75 mm]{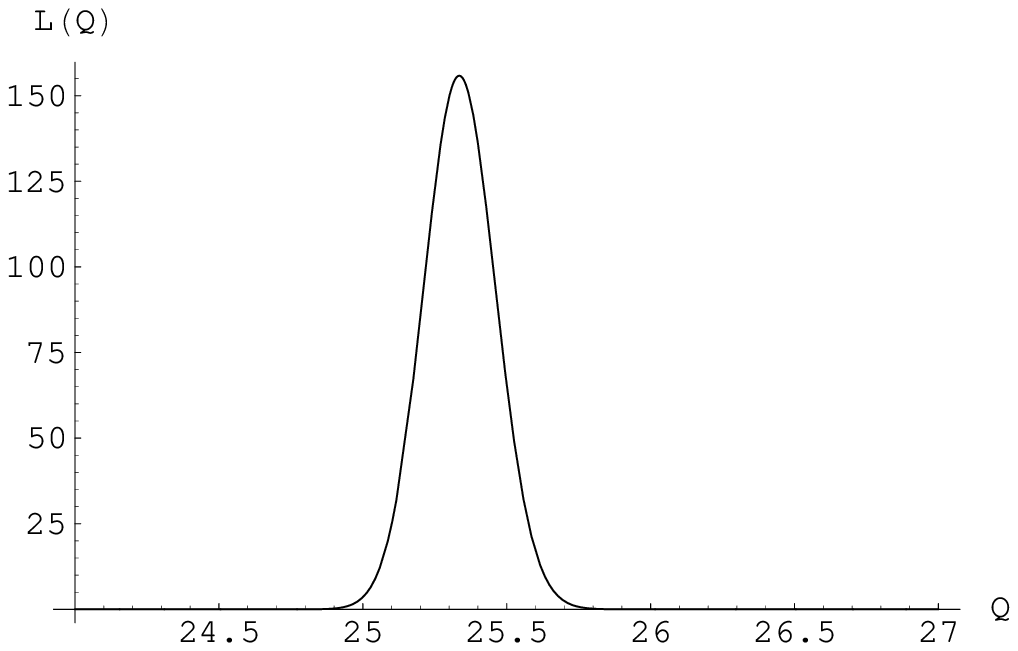}
\caption{Individual likelihood functions (\ref{lf0}) for the light
elements.  From top to bottom: D, $^4$He (Eq.\ref{hehigh}), Li, and
the combined datset.}
\end{figure}

From the deuterium likelihood function we found the position of the
maximum and 1$\sigma$ deviations:
\begin{equation} \label{dmax}
 Q_D=  (25.74 + 0.92 - 0.68)\times 10^9.
\end{equation}
The shape near the maximum is apparently non-symmetric. The position of the 
maximum is fully compatible with the present value of $Q=25.82\times 10^9$ K.  
The helium likelihood function is much narrower (see the second panel from 
the top). It gives for 
the maximum and for the 1$\sigma$ the values
\begin{equation} \label{hemax}
Q_{He}=(25.37 \pm 0.13) \times 10^9.
\end{equation}
This value lies  below the present value of the binding
energy. Finally, the lithium likelihood function has the maximum at
\begin{equation} \label{limax}
Q_{Li}=(24.88 + 0.43 -0.59) \times 10^9.
\end{equation}
The position of this maximum is compatible with the helium result.

The general likelihood function (\ref{lf0})  is plotted in the lower panel in 
Fig. 4
The position of its maximum differs only slightly  from the position given by
the helium likelihood function.
\begin{equation}
Q_{BBN} = (25.34 \pm 0.12)\times 10^9
\end{equation}

It is interesting to compare the light element abundances for two values of the
deuterium binding energies. In Fig. 5 we plotted the traditional curves 
for the light element
abundances as a functions of $\eta$ for two values of $Q$. The dotted lines in
the figures correspond to a present value of $Q_{present}=25.82\times 10^9$ K, 
while the solid curves
correspond to a new value $Q_{BBN}=25.34\times 10^9$ K. Clearly, the new value
$Q_{BBN}$ moves the curves closer to the data.
\begin{figure}
\includegraphics[width=7.5cm]{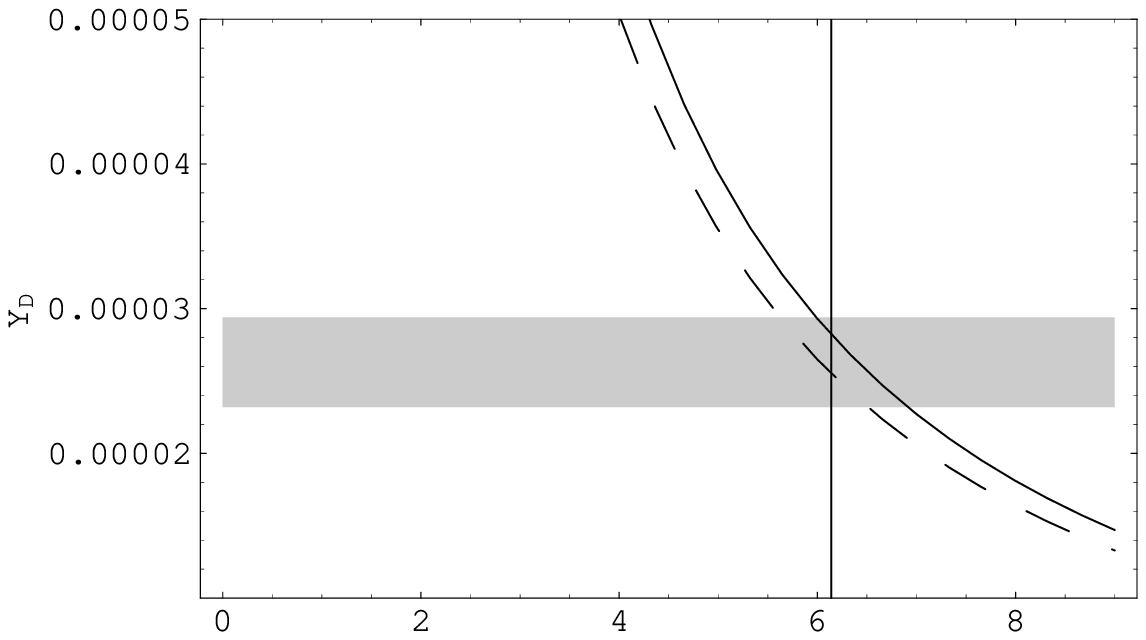}
\includegraphics[width=7.5cm]{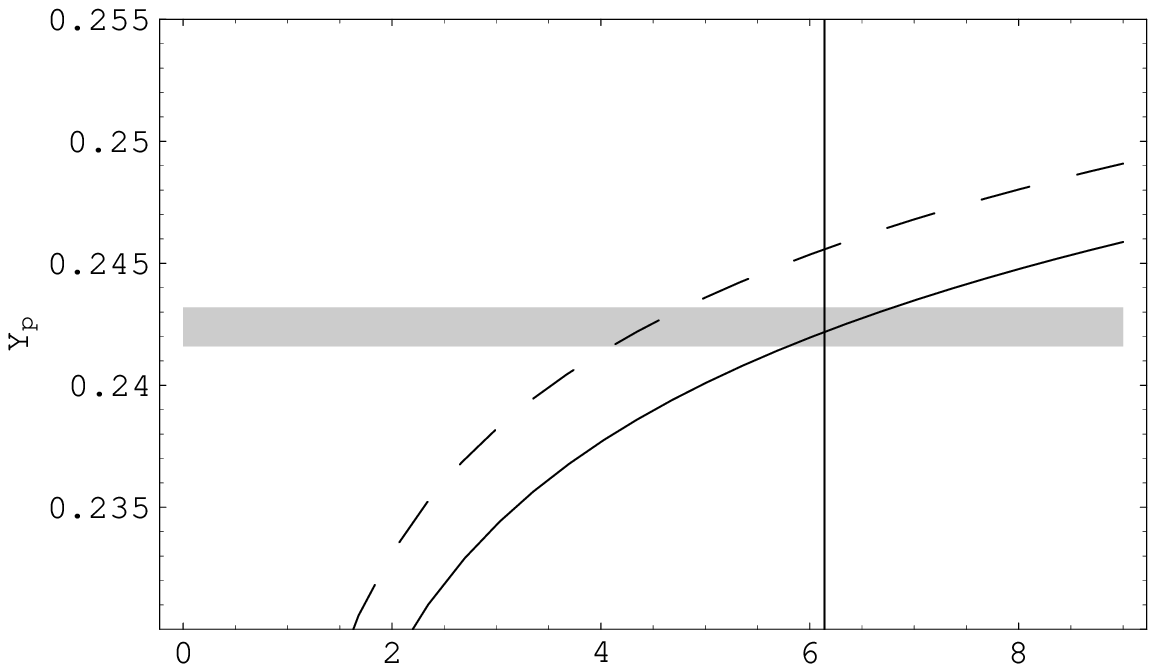}
\includegraphics[width=7.5cm]{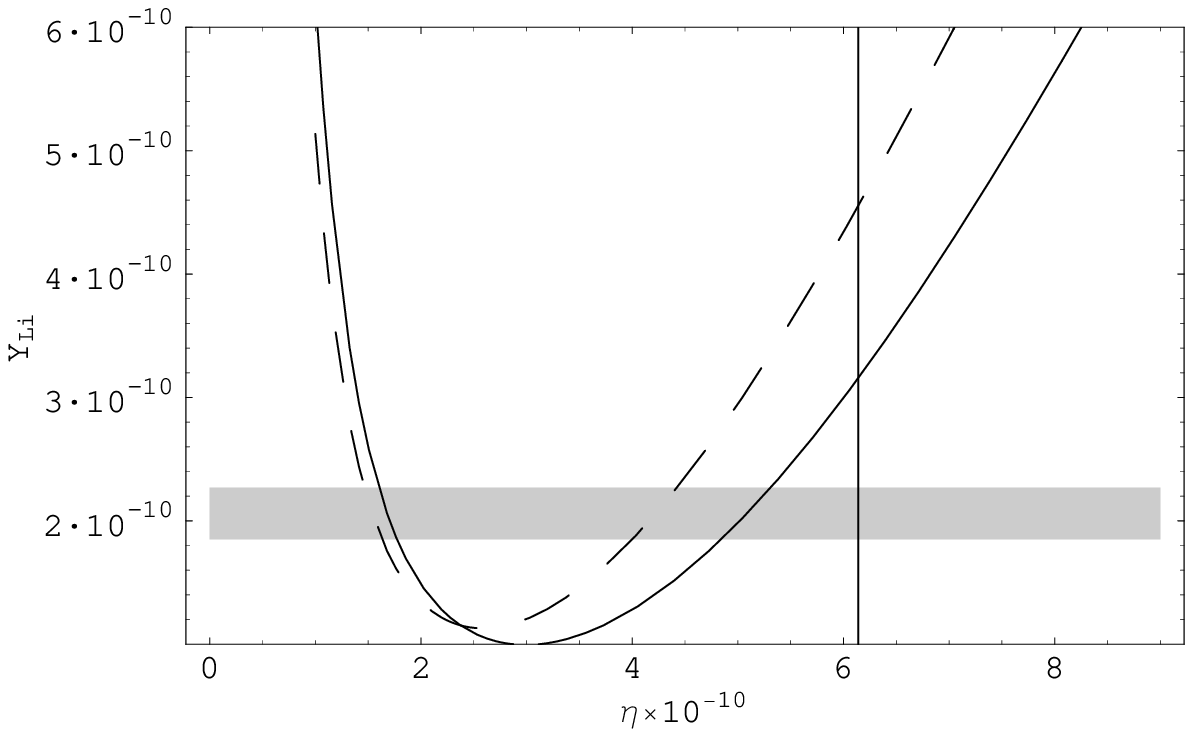}
\caption{The predicted light element abundance yields as a function of
$\eta$, for two values of the deuteron binding energy, $Q$. The dotted
curve corresponds to the present value of $Q_{present}=25.82\times
10^9$ K. The solid curve corresponds to the new value of $Q=Q_{BBN} =
25.34\times 10^9 $ K. The vertical line corresponds to $\eta=6.14$
(WMAP value). The shaded regions is the 1$\sigma$-ranges for the
observed light element abundances, where Y$_p$ is from
eq.\ref{hehigh}.}
\end{figure}

The result which we obtained may be presented as
\begin{equation}\label{Qhigh}
\delta Q/Q = -0.019 \pm 0.005,
\end{equation}
where $\delta Q=Q_{BBN}-Q_{present}$.
If we do not fix $\eta$ and try to fit it simulaneously with $Q$ we obtain
\begin{equation}\label{Qetahigh}
\delta Q/Q = -0.022 \pm 0.008,\;\eta = (6.51 +0.77-0.66)\times 10^{-10}.
\end{equation}
The obtained $\eta$ is fully compatible with the one measured by WMAP.  

These values of $\delta Q/Q$ and $\eta$ were obtained for high value of the 
helium mass fraction $Y_p$. If we use as an input the low value of $Y_p$ from 
(\ref{helow})
we obtain
\begin{equation}\label{Qlow}
\delta Q/Q = -0.048 \pm 0.004,
\end{equation}
If we fit both $\delta Q/Q$ and $\eta$ we 
obtain
\begin{equation}\label{Qetalow}
\delta Q/Q = -0.059 \pm 0.007,\;\eta = (7.55 +0.91-0.75)\times 10^{-10}.
\end{equation}
Finally if we use the value of $Y_p$ for $^4$He obtained using the whole sample
of 14  points, with increased error bars, from
Eq.(\ref{hedata}), we obtain
\begin{equation}\label{Qlim}
\delta Q/Q = -0.033 \pm 0.006,
\end{equation}
and for $\delta Q/Q$ and $\eta$
\begin{equation}\label{Qetalim}
\delta Q/Q = -0.042 \pm 0.009,\;\eta = (7.00 +0.85-0.72)\times 10^{-10}.
\end{equation}
The results given in eqs.(\ref{Qhigh}) and (\ref{Qlow}) therefore
represent an estimate of the plausible range in $\delta Q/Q$.  Despite
the clear systematic uncertainties in the $^4$He data, and accepting
the WMAP value of $\eta$ as being correct, $\delta Q/Q$ appears to
deviate from zero by 4$\sigma$ (eq. \ref{Qhigh}) or greater
(eqs. \ref{Qlow}, \ref{Qlim}).

The deuteron binding energy depends on the strong scale and quark
masses. It is convenient to assume that $\Lambda_{QCD}$ is constant ,
and the quark mass is variable.  This only means that we measure all
energies in units of $\Lambda_{QCD}$ (and cross-sections in units
$\Lambda_{QCD}^{-2}$).  In Ref. \cite{FS1} we concluded that the
deuteron binding energy is very sensitive to variation of the strange
quark mass $m_s$ \cite{strange}:
\begin{equation}\label{Qms}
\frac{\delta (Q/\Lambda_{QCD})}{(Q/\Lambda_{QCD})} = -17
 \frac{\delta (m_s/\Lambda_{QCD})}{(m_s/\Lambda_{QCD})}
\end{equation}
Combining eqs. (\ref{Qlim}) and (\ref{Qms}) we obtain
\begin{equation}
 \frac{\delta (m_s/\Lambda_{QCD})}{(m_s/\Lambda_{QCD})}=
 (1.1 \pm 0.3)\times 10^{-3}
\end{equation}
This equation may contain an additional factor (close to one)
reflecting unknown theoretical uncertainty in eq. (\ref{Qms}).  Note
that we obtain here variation at the level $10^{-3}$ while the limits
on variation of $\alpha$ \cite{Bergstrom,uzan} and
$\Lambda_{QCD}/M_{Plank}$ \cite{FS,uzan} are an order of magnitude
weaker. This may serve as a justification of our approach.

\section{Conclusion}
Allowing the deuteron binding energy, $Q$, to vary in BBN appears to
provide a better fit to the observational light element abundance
data.  Varying $Q$ simultaneously does two things; it resolves the
internal inconsistency between $^4$He and the other light elements,
and it also results in excellent independent agreement with the baryon
to photon ratio determined from WMAP.  (Fig. 5). However, the
magnitude of the variation is sensitive primarily to the observed
$^4$He abundance, which has the smallest relative statistical error.
A systematic error in the abundance of $^4$He could imitate the effect
of the deuteron binding energy variation, although one needs a
systematic error which is very much greater than has been claimed in
the most recent observational work.

We note that Izotov and Thuan \cite{izth03}, the most recent estimate
for Y$_p$ in our sample, argue that systematics are at most 0.6\% for
that survey.  On the other hand, the possibility has also been
explored that the creation of $^4$He in population III stars might
mean that the true primordial $^4$He abundance is lower even than that
seen in the most metal-poor objects \cite{salv}.  If so, the
significance of the deviation of $\delta Q/Q$ from zero we report in
this paper would be even larger.

These results hopefully provide an extremely strong motivation to
obtain substantially better measurements of all the light elements,
and to explore even more intensively, the possible sources of
systematic errors.

\acknowledgments {This work was supported by the Australian Research
Council and Gordon Godfrey fund.  We are also grateful to the John
Templeton Foundation for support. We thank G. Steigman and J. Barrow for
informative discussions.}

\appendix

\section*{Appendix}
Let $U_0(r)$ be a critical depth potential for which the binding energy $Q=0$, 
and $U_t(r)$ a potential for a proton neutron system in a triplet state 
producing a deuteron with small binding energy $Q=2.22$ MeV. If we add to
the deuteron Hamiltonian a perturbation 
$$
\delta U_\lambda(r)=\lambda(U_0(r) - U_t(r)), 
$$
then, variation of  $\lambda$ from 0 to 1 will move the binding energy $Q$ 
from 2.22 MeV to 0.
From a virial theorem for a quantum system we have 
\begin{equation} \label{ben}
\frac{dE}{d\lambda} = \int_0^\infty(U_0(r)-U_t(r))\chi^2(r)dr, 
\end{equation}
where
$\chi(r)$ is the radial s-wave function. For simplicity we neglect the d-wave 
contribution. For $Q\rightarrow 0$ the main contribution into normalization 
integral for $\chi(r)$ comes from the region outside of the nuclear forces 
radius
$R$. The normalization integral can be presented as sum of contributions from 
inner and outer regions
\begin{equation}\label{norm}
\int_0^R \chi^2(r) dr + b^2\int_R^\infty e^{-2\gamma r} dr=1,
\end{equation}
where $\gamma=\sqrt{\frac{m_p|E|}{\hbar^2}}$. At $|E|\rightarrow 0$ the second 
integral dominates giving $b^2=2\gamma$. Separating the $E$-dependence of the 
normalization factor in $\chi(r)$ we can rewrite Eq.(\ref{ben}) as
\begin{equation}\label{ben1}
\frac{dE}{\sqrt{|E|}} =d\lambda\, 2\sqrt{\frac{m_p}{\hbar^2}}\int_0^\infty
(U_0(r)-U_t(r))\tilde{\chi}^2(r)dr,
\end{equation}
where $\tilde{\chi}(r)$ is practically independent on $E$ inside the potential
well (where $E<<U$) and $\tilde{\chi}(r)=e^{-2\gamma r} 
\rightarrow 1$ at $r>R$ when $|E|\rightarrow 0$ . Integrating the left hand 
side of Eq.(\ref{ben1}) over $E$ from $-Q$ to 0 and the right hand side of 
 Eq.(\ref{ben1}) over $\lambda$ from 0 to 1 we obtain
\begin{equation}\label{ben2}
Q = \frac{m_p}{\hbar^2}\left(\int_0^\infty(U_0(r)-U_t(r))\tilde{\chi}^2(r)dr\right)^2. 
\end{equation}
The Eq.(\ref{ben2}) shows that the position of a shallow bound level depends
quadratically on the difference between the actual depth of the potential and 
the critical one. For a square well $Q=\frac{\pi^2 (U-U_0)^2}{16 U_0}$,
$U_0 = \frac{\pi^2 \hbar ^2}{4 m_p R^2}$.

In fact, the Eq.(\ref{ben2}) is valid not only for the energy of a bound level
but for the energy of a virtual level as well.
The integration in Eq.(\ref{ben2}) is over the region $r<R$ where the function
$\tilde{\chi}^2(r)$ is insensitive to the energy $E$, and the quadratic 
dependence on $U-U_0$ guarantees the validity of Eq.(\ref{ben2}) for both  
$U<U_0$ and $U>U_0$. Thus, for the energy of the virtual level we have
\begin{equation}\label{ven}
\epsilon_v = \frac{m_p}{\hbar^2}\left(\int_0^\infty(U_0(r)-U_s(r))
\tilde{\chi}^2(r)dr\right)^2,
\end{equation}
where $U_s(r)$ is the potential for a singlet states. We have both $Q<<U$ and
$\epsilon_v << U$. This means that the difference between  the triplet
 and  singlet potentials is not large. Assuming  that the
changes in the triplet and  singlet potentials are the same we obtain 
for the changes in $Q$ and $\epsilon_v$ the relation
\begin{equation}\label{rel}
\frac{\delta \epsilon_v}{\sqrt{\epsilon_v}}=-\frac{\delta Q}{\sqrt{Q}}.
\end{equation}
This equation also holds for the effect produced by variation of the proton
 mass ( the dominating effect comes from variation of $U_0$).


\begin{thebibliography}{99}
\bibitem{strange} The enhanced sensitivity of the deuteron binding
energy to the strong qurak mass is explained by four factors \cite{FS1}:
1) deuteron is the shallow level which may be eliminated by small
variation of the binding potential; 2) there is strong cancellation
between $\sigma$-meson and $\omega$-meson contributions into nucleon-nucleon
interaction (Walecka model), therefore, a minor variation
of $\sigma$-meson mass  leads to a significant change in the strong
potential; 3) $\sigma$-meson contains valence s- and anti-s quarks
 which give large contribution to its mass; 4) $\sigma$-meson
energy level is repelled down due to mixing with  close $K \overline{K}$
 state. Some strange mass contribution also comes from the proton mass.
\end{thebibliography}
\end{document}